\documentclass[conference,a4paper]{IEEEtran}
\usepackage[latin9]{inputenc}
\usepackage{amsmath}
\usepackage{amssymb}
\usepackage{graphicx}

\usepackage{cite}

\IEEEsettopmargin{t}{60pt}

\begin{document}
\title{Wireless LAN sensing with smart antennas}
\author{\IEEEauthorblockN{Marco Santoboni\IEEEauthorrefmark{1}\IEEEauthorrefmark{2}, 
Riccardo Bersan\IEEEauthorrefmark{1}, 
Stefano Savazzi\IEEEauthorrefmark{3}, 
Alberto Zecchin\IEEEauthorrefmark{1}, 
Vittorio Rampa\IEEEauthorrefmark{3} 
Daniele Piazza\IEEEauthorrefmark{1} 
} 
\IEEEauthorblockA{\IEEEauthorrefmark{1}
Adant Technologies Inc. C.so Stati Uniti, 35127 Padova, Italy, e-mail:
\{name.surname\}@adant.com} \IEEEauthorblockA{\IEEEauthorrefmark{2}
Università di Padova, Padova, Italy} \IEEEauthorblockA{\IEEEauthorrefmark{3}Consiglio Nazionale delle Ricerche (CNR), IEIIT
institute, Milano, Italy, e-mail:\{name.surname\}@ieiit.cnr.it}}
\maketitle
\begin{abstract}
The paper targets the problem of human motion detection using Wireless
Local Area Network devices (WiFi) equipped with pattern reconfigurable
antennas. Motion sensing is obtained by monitoring the body-induced
alterations of the ambient WiFi signals originated from smart antennas
supporting the beam-steering technology, thus allowing to channelize
the antenna radiation pattern to pre-defined spots of interest. We
first discuss signal and Channel State Information (CSI) processing
and sanitization. Next, we describe the motion detection algorithm
based on Angle-of-Arrival (AoA) monitoring. Proposed algorithms are
validated experimentally inside a large size smart home environment. 
\end{abstract}

\vskip0.5\baselineskip 

\section{Introduction}

Wireless Local Area Network (WLAN) sensing (also known as WiFi sensing)
targets the integration of novel environmental recognition capabilities
into next generation WiFi and cellular machine-type communication
radio interfaces \cite{mustafa,patwari1,mag2017}. The Channel State
Information (CSI) of ambient Wi-Fi signals, optimized for communication,
are processed in real time to detect changes of the environment, such
as human body or object motions, activities, gestures as well as bio
metric measurements \cite{sleep}. Radio signals are perturbed by
objects, body movements, and changes in the surroundings, as a result
of the propagation of ElectroMagnetic (EM) waves \cite{mag2019}.
Hence, in addition to transporting modulated information, WiFi signals
can be re-used for sounding the environment in the form of a 2D/3D
views of traversed objects by the EM wavefield.

Presently, the joint utilization of radio resources for communication
and sensing, is actively discussed in research, standardization and
industry. Several methods for passive radio sensing have been studied
\cite{patwari1,mag2017,mag2019} based on simple radio-frequency (RF)
hardware and power only measurements in the form of Received Signal
Strength (RSS) values. Recently, more complex methods based on Channel
State Information (CSI) matched with multi-antenna orthogonal frequency
division multiplexing (MIMO-OFDM) devices have been shown to provide
improved detection and recognition accuracy \cite{csi,defi,MIMO}.
Antenna arrays can be also exploited \cite{music} to extract Angle-of-Arrival
(AoA) information (\emph{i.e.}, beam-forming) and improve both spatial
accuracy and resolution. 

\begin{figure}
	\centering
	\includegraphics[width=0.9\columnwidth]{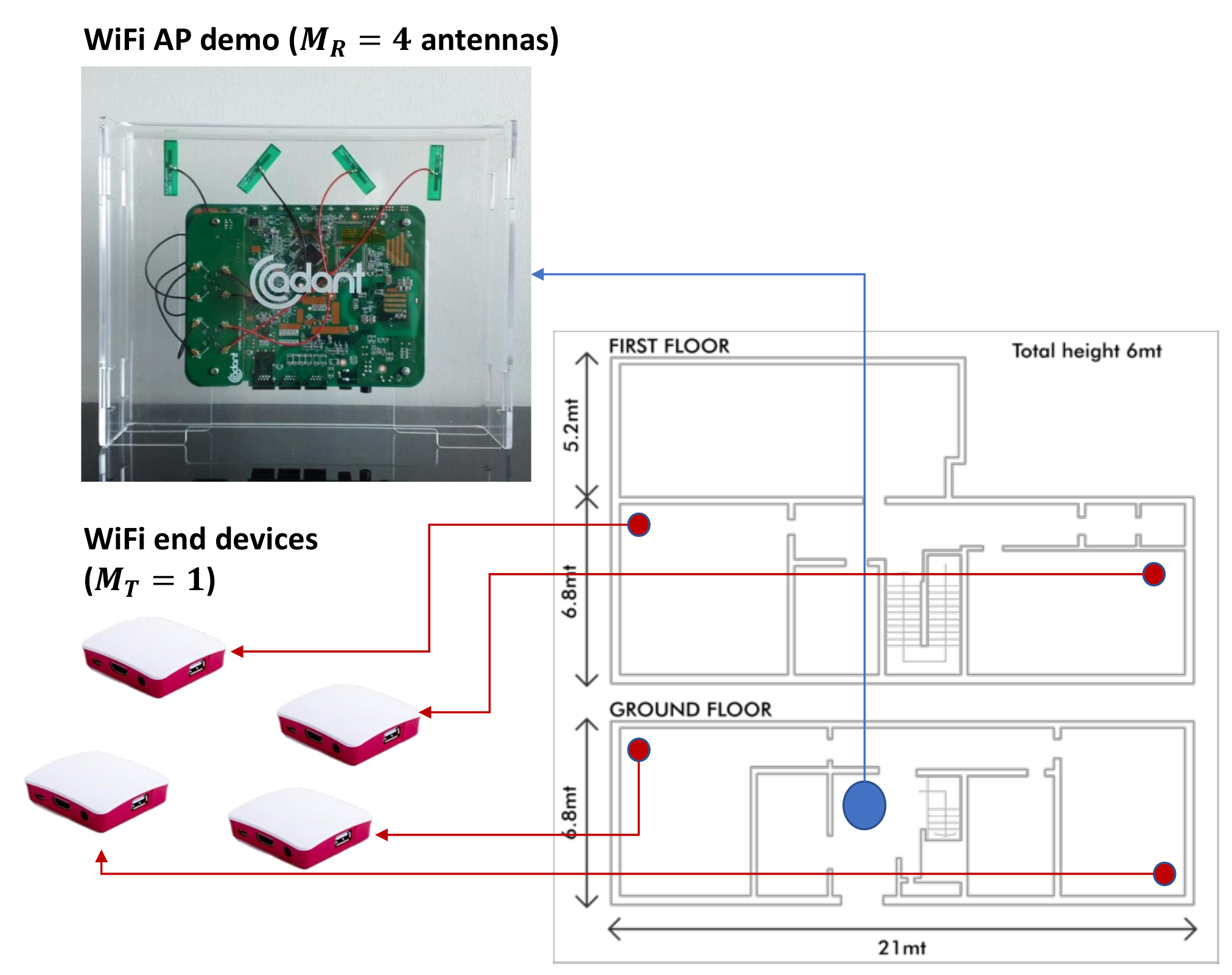}
	\caption{WiFi access point (AP) equipped with $M_{R}=4$ smart antenna array and deployment setup for motion sensing tests: 4 WiFi nodes (end devices) with single antennas ($M_{T}=1$) and floor map of network layout inside a residential home (with concrete walls). The test environment has size 340 sqm and characterized by concrete walls.}
	\label{intro}
\end{figure}

Motion and intrusion detection are the most used and studied application
of WLAN sensing. The main approaches towards radio sensing are both
data and model driven as human behaviors and movements can be modelled
through a training stage \cite{15,16} or using statistical/EM tools
\cite{tap,mag2019}. Body motions cause abnormalities in radio propagation
and can be detected using several methods. For example, model-based
algorithms, \emph{e.g.}, threshold-based detection, and simple machine
learning based algorithms, \emph{e.g.}, Support Vector Machine (SVM),
are widely used. First studies on baseband CSI processing dates back
to \cite{14}: the DeMan system is a unified scheme for non-invasive
detection of moving and stationary human on commodity WiFi devices.
DeMan takes advantages of both amplitude and phase eigenvalues information
of CSI to detect moving targets and provides a detection accuracy
of about $95$\% for both moving and stationary people moving in controlled
indoor environments. MoSense \cite{15} is a radio frequency (RF)
based device-free motion detection system designed to deliver a reliable
and transparent detection service in the real-time. Designed for MIMO-OFDM
radio interfaces, the implementation of MoSense tackles the problem
of optimal selection of CSI subcarriers, to better capture the impact
of motions from the noisy channel samples. The features used for motion
detection are the phase difference from CSI signals received across
two antennas, which lead to an accuracy of $97$\%. Hang et al. in
\cite{16} designed and implemented WiSH, a real-time system for contactless
human detection that can be implemented on resource limited devices
with limited computational power and low CSI sampling rate (20 Hz).


The integration of radio sensing algorithms with smart-antenna systems
is a new paradigm that will become a key enabling technology for next
generation environment-aware radio communication system. Building
on the results from initial studies in \cite{pimrc}, the paper addresses
the technology transfer and validation of radio sensing as integrated
within a WiFi radio interface compliant with the IEEE 802.11ax (WiFi6)
standard and equipped with smart antennas supporting the beam-steering
technology \cite{reconfigurable}. Pattern reconfigurable antennas
allow to channelize the antenna radiation pattern to pre-defined areas
of interest \cite{pimrc,reconfigurable}. The paper proposes CSI signal
pre--processing tools required to effectively transform beam-steering
into a reliable sensing technology. In fact, adapting the radiation
pattern of individual antennas provides more spatial views of the
environment and hence promises unprecedented sensing accuracy as well
as it paves the way to joint sensing and communication co-designs.
In particular, we propose a beam-space processing model and an AoA
estimation framework based on Principal Component Analysis (PCA) and
MUltiple SIgnal Classification (MUSIC). Several validation tests have
been carried out inside a smart home environment, namely a large test
house of approximately 300 sqm (see Fig. \ref{intro}), with the purpose
of: a) developing an enhanced WiFi Access Point (AP) gateway for residential
use with improved connectivity, and b) providing, through the same
residential WiFi test network, a motion detection service to support
several application verticals such as intrusion detection and smart
living.

The paper is organized as follows. Sect. II describes the CSI modelling
and processing of AoA including the necessary adaptations to the proposed
smart antenna system as well as the main implementation choices for
AoA estimation, CSI feature processing and motion detection. Sect.
III describes the case study and the results obtained in the test
house environment: according to this test setup, system performances,
namely accuracy, false alarm,miss detection probability and robustness,
are verified considering two AP deployment cases. Sect. IV draws some
conclusions.

\section{CSI modelling and AoA processing}

We analyze the uplink of a typical WiFi MIMO-OFDM communication system
\cite{MIMO,WiFi} consisting of $N$ radio devices (WiFi6 end devices)
equipped with $M_{T}$ antennas and communicating with one WiFi access
point (AP) gateway equipped with $M_{R}$ antennas with spacing $D$.
In line with \cite{pimrc}, we resort to a practical case where the
AP gateway is equipped with a smart antenna system and supports beam-steering
functions \cite{patent}, while the WiFi6 end devices are equipped
with a single antenna, $M_{T}=1$. We consider the problem of device-free
target motion detection inside an indoor environment consisting of
multiple rooms and different subjects. Body motions are detected by
inspection and real-time analysis of the CSI response $\mathbf{H}(t)$
that is observed at discrete time instants $t=1,2,...,$ of consecutive
received WiFi PHY Protocol Data Unit (PPDU) frames.

\begin{figure}
	\centering
	\includegraphics[width=0.8\columnwidth]{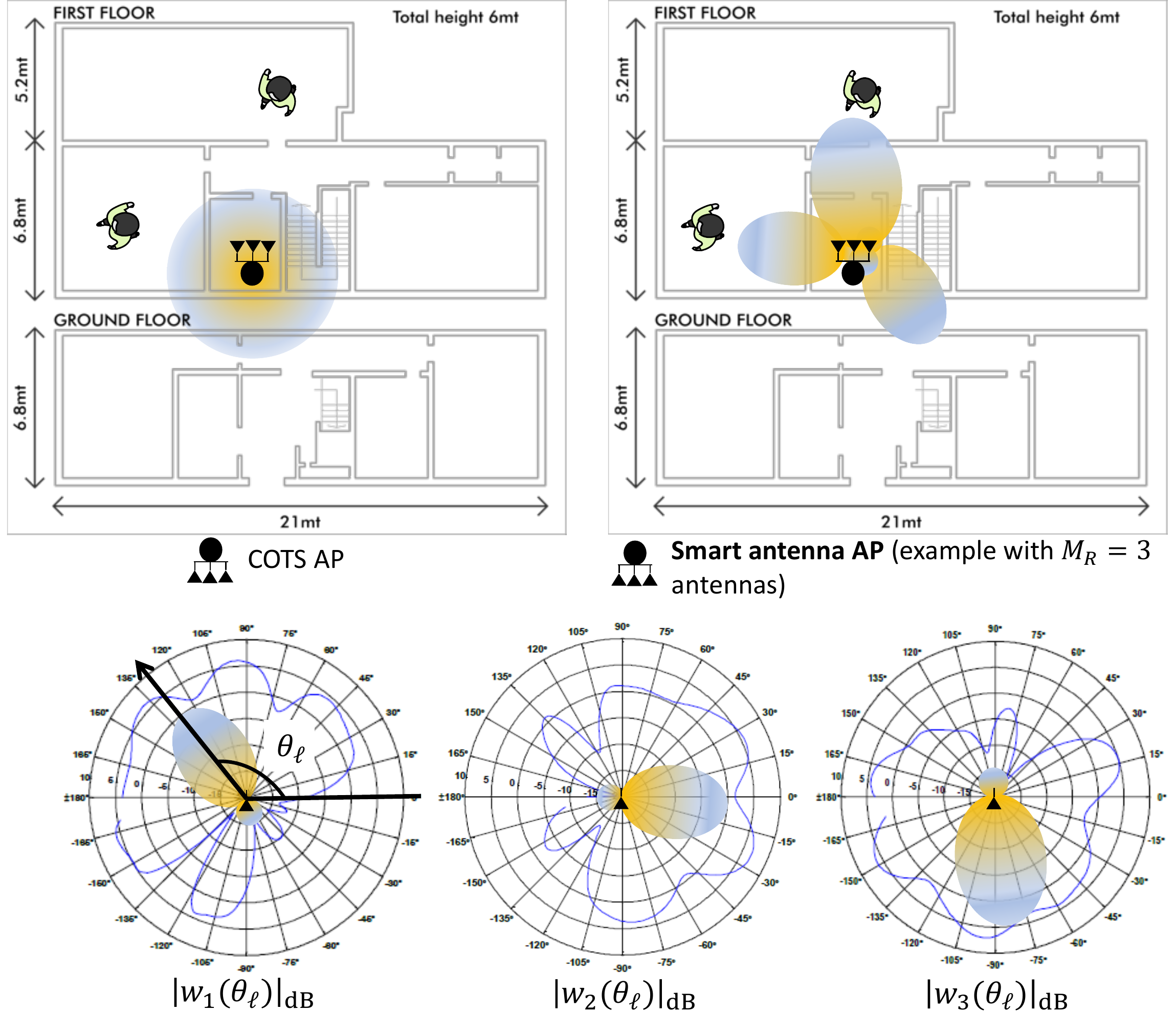}
	\caption{WiFi sensing with smart antennas (beam-steering antennas) compared with COTS antenna devices. Bottom: examples of beampatterns.}
	\label{beams}
\end{figure}

The WiFi signal from the end device typically reflects off multiple
objects (around $6\div8$ indoor reflectors\cite{music_gen}) when
approaching the AP. Considering that the presence of the subject moving
in the environment modifies the AoA paths, and that the number of
antennas at the AP are limited, the MUSIC algorithm is typically adopted
\cite{music,mmusic} to disentangle the multipath components and estimate
the AoA of the dominant propagation paths. The received CSI vector
$\mathbf{h}_{s}$ at the AP antenna array over one pilot subcarrier
$s$, with carrier frequency $f_{s}$, is generally obtained by the
superposition of signals due to all the paths. Considering an OFDM
frame structure with $S$ pilots ($s=1,...,S$), $L$ dominant reflections/propagation
paths ($\ell=1,...,L$) with AoA $\theta_{\ell}$ and the corresponding
complex attenuation $\gamma_{\ell,s}$, it is 
\begin{equation}
\mathbf{h}_{s}=\sum_{\ell=1}^{L}\mathrm{\mathbf{w}_{\ell}}\odot\mathbf{a}(\theta_{\ell})\cdot\gamma_{\ell,s}\label{eq:hs}
\end{equation}
where the steering vectors $\mathbf{a}(\theta_{\ell})=\left\{ \exp\left[-jm\varphi(\theta_{\ell})\right]\right\} _{m=0}^{M_{R}-1}$
model the phase shifts $\varphi(\theta_{\ell})=2\pi D\sin(\theta_{\ell})f_{s}/c$
at the antenna array \cite{mmusic} ($c$ is the light speed), while
$\odot$ is the Hadamard (element-wise) product with the beam pattern
vectors $\mathbf{w_{\ell}}=w_{\ell}(\theta_{\ell})=[w_{1}(\theta_{\ell}),...,w_{M_{R}}(\theta_{\ell})]^{T}\in\mathbb{C}^{M_{R}\times1}$
that collects the corresponding complex valued antenna responses for
the considered AoA $\theta_{\ell}$, according to the chosen antenna
beam pattern. Real beampattern examples are shown in Fig. \ref{beams}.
Assuming that the antenna response patterns do not change across closely
spaced subcarriers \cite{music,music_gen}, the CSI matrix $\mathbf{H}_{t}\in\mathbb{C}^{M_{R}\times S}$
at time $t$ is 
\begin{equation}
\mathbf{H}(t)=\left[\mathbf{h}_{1},...,\mathbf{h}_{S}\right]=\left(\mathrm{\mathbf{W}\odot}\mathbf{S}\right)\boldsymbol{\gamma}(t)\label{eq:hs-1}
\end{equation}
with $\mathbf{S}=[\mathbf{a}(\theta_{1}),...,\mathbf{a}(\theta_{L})]$,
and, similarly, the beam pattern terms $\mathbf{W}=[\mathbf{w}_{1},...,\mathbf{w}_{L}]$.
Finally $\boldsymbol{\gamma}(t)=\left[\gamma_{\ell,s}\right]\in\mathbb{C}^{L\times S}$
contains the corresponding complex fading attenuations at time $t$:
notice that we assume that AoAs and beam patterns are stationary for
a number $T$ of consecutive PPDU frames.

Considering now the CSI $\mathbf{Y}(t)\in\mathbb{C}^{M_{R}\times S}$
that is measured/observed by the receiver antennas at time $t$, 
\begin{equation}
\mathbf{Y}(t)=\mathbf{H}(t)+\mathbf{N}(t),\label{eq:mod}
\end{equation}
with the superimposed AWGN noise $\mathbf{N}_{t}$, a PCA-based algorithm
is adopted for the eigenstructure analysis of the correlation matrix
$\mathbf{R}_{\mathbf{Y}}=\mathbb{E}_{t}\left[\mathbf{Y}\cdot\mathbf{Y}^{H}\right]\in\mathbb{R}^{M_{R}\times M_{R}}$.
Keeping into account the beam-steering process, we define $\mathbf{\widetilde{S}}=\mathrm{\mathbf{W}\odot}\mathbf{S}=\left[\mathbf{\widetilde{a}}(\theta_{1}),...,\mathbf{\widetilde{a}}(\theta_{L})\right]$
with $\mathbf{\widetilde{a}}(\theta_{\ell})=\mathrm{\mathrm{diag}(\mathbf{w}_{\ell})}\cdot\mathbf{a}(\theta_{\ell})$,
so that 
\begin{equation}
\mathbf{R}_{\mathbf{Y}}=\mathbf{\widetilde{S}}\mathbf{R}_{\boldsymbol{\gamma}}\mathbf{\widetilde{S}}^{H}+\sigma^{2}\mathbf{I},
\end{equation}
where $\mathbf{R}_{\boldsymbol{\gamma}}$ is defined as $\mathbf{R}_{\boldsymbol{\gamma}}=\mathbb{E}_{t}\left[\boldsymbol{\gamma}\cdot\boldsymbol{\gamma}^{H}\right]$
and $\sigma^{2}$ is the AWGN noise power. The $L<M_{R}$ dominant
AoAs (\emph{i.e.}, corresponding to the signal subspace) correspond
to the highest peaks of the so-called spatial spectrum \cite{music}\footnote{We assume that the vectors $\mathbf{\widetilde{a}}(\theta_{\ell})$
are still mutually orthogonal and serve as basis for the PCA signal
subspace, as required by the PCA-based analysis.} 
\begin{equation}
\widehat{\theta}\simeq\arg\max_{\theta}\frac{1}{\mathbf{a}^{H}(\theta)\cdot\mathbf{E}_{m}\mathbf{E}_{m}^{H}\cdot\mathbf{a}(\theta)},\label{eq:mu}
\end{equation}
with $\mathbf{E}_{m}$ the $m=M_{R}-L$ eigenvectors of $\mathbf{R}_{\mathbf{Y}}$
corresponding to the $m$ smallest eigenvalues (\emph{i.e.}, the noise
subspace). Notice that $\mathbf{R}_{\mathbf{Y}}$ is estimated by
maximum likelihood over $T$ consecutive PPDU frames. The resulting
$L$ dominant components in (\ref{intro}) might be generally different
from the true ones as the result of the beamsteering process, as $\mathbf{\widetilde{a}}(\theta_{\ell})\neq\mathbf{a}(\theta)$;
however, in what follows, we show that they can be used as an effective
subspace basis for motion detection. In the example of Fig. \ref{intro},
an AP device equipped with $M_{R}=4$ electronically steerable antennas
measures the PHY layer CSI over $S=53$ pilot subcarriers: the correlation
matrix $\mathbf{R}_{\mathbf{Y}}$ is obtained from $T=7$ consecutive
frames, corresponding to $250$ms interval. Frames are transmitted
by an end device device (Raspberry PI 4) equipped with standard, vertically
polarized omnidirectional antenna. For the purpose of motion detection,
we detect and track $L=2$ dominant components.

\subsection{Phase information pre-processing and sanitization}

Real phase measurements are affected by many key factors, of which
the most important are the Sampling Time Offset (STO) and the Sampling
Frequency Offset (SFO). These factors make raw CSI measurements on
commercial WiFi devices extremely noisy. For example, the STO brings
out extra delay in addition to the signal propagation time of the
interested environment. This additional time delay results in random
phase offset that affects the CSI readings, which in turn contaminates
the true phase shifts. The SFO affects the sampling time offset from
packet to packet thus adding noise to the phase estimates across packets.
Therefore, it is necessary to clean up raw phase measurements before
performing the AoA estimation. The CSI phase pre-processing is divided
in two steps: unwrap and sanitization.

The first step is the unwrapping of the CSI phase measurement, which
comes out wrapped in the interval $[-\pi,\pi]$ from the AP CSI extractor
tool: the unwrapping process corresponds to the correction of the
actual phase by a multiple of $2\pi$ \cite{defi}. STO adds a constant
offset to the Time of Flight (ToF) estimates of all paths: this additional
delay manifests itself as a linear frequency term in the phase response
of the channel. The random phase offset at the $s$-th subcarrier
is therefore $2\pi(s-1)f_{\Delta}\tau_{s}$, where $\tau_{s}$ denotes
the time delay due to the STO and $f_{\Delta}=312.5$ KHz represents
the frequency interval between the adjacent subcarriers. According
to \cite{defi}, a linear regression is adopted to estimate the STO
and random phase offset from the unwrapped phase $\phi_{s,m}(t)=\left\{ \angle\mathbf{Y}(t)\right\} _{s,m}$
observed on the subcarrier $s$ and the antenna $m$ as
\begin{equation}
\left(\widehat{\tau}_{s},\widehat{\xi}\right)_{i}=\arg\min_{\tau_{s},\xi}\sum_{s,m}\left[\phi_{s,m}(t)+2\pi(s-1)f_{\Delta}\tau_{s}+\xi\right]^{2}.
\end{equation}
The calibrated phase response can be corrected (sanitized) as $\widehat{\phi}_{s,m}(t)=\phi_{s,m}(t)-2\pi(s-1)f_{\Delta}\widehat{\tau}_{s}-\widehat{\xi}$.
A weighted moving average has been also applied on received CSI amplitude
$\left|\mathbf{Y}(t)\right|$ considering $T=7$ consecutive frames.


\subsection{AoA based motion detection}

The approach we followed is to detect and classify the variations
of the estimated dominant signals AoA (\ref{eq:mu}) as these are
indicative of the subject presence as moving in the environment. In
case of motion, CSI perturbations might affect one or all the dominant
AoAs: in line with EM modelling of body-induced fading \cite{tap},
this perturbation is higher when the movement occurs close to the
WiFi end device or the AP. After the CSI pre-processing and phase
sanitization, we apply the AoA estimation using the PCA-based method
previously described. Considering the beam-steering process, the estimated
dominant AoAs are not coherent with the real direction of the signals.
In order to solve this issue, a phase calibration is implemented during
the initial system warm-up, which correct the phase differences among
each antenna element. Phase calibration is a one-time process and
must be carried out before every installation. Another limitation
is that when the movement happen behind a sensor, it barely affects
the AoAs, which can lead to a missed alarm event. For such cases,
optimal deployment of WiFi end devices is advised. In the following
tests, we extract and monitor $L=2$ dominant AoAs (although up to
$3$ paths might be resolvable) to maximize the system robustness:
these monitored angles are always detected, or with high probability,
as they typically correspond to the main path, \emph{i.e.,} the Line-of-Sight
(LOS) of near LOS path, and the main reflected path (\emph{e.g.},
caused by relevant obstacles). Finally, AoA resolution improvements
are implemented according to \cite{mmusic} which remove the coherence
among signals.


\subsection{CSI features for motion detection}

\label{sec: beam}

Besides the AoA estimates, other signal features can be extracted
from the CSI and the Received Signal Strength (RSS) data \cite{MIMO}.
In particular, we focus below on several RSS statistical metrics (average
and standard deviation) evaluated over consecutive WiFi frames, the
time-frequency correlation of CSI matrix and a measure of CSI phase
changes. All the features have been extracted and analyzed in order
to understand which of them are more sensible and reliable when a
motion event occurs. The proposed decision process monitors both the
dominant AoAs, as described in the previous section, and the above
mentioned metrics. The approach we followed is to select the best
pool of features for our decision process, and analyse advantages,
disadvantages and limitations of these choices. Notice that all the
proposed features compare the CSI/RSS samples over two adjacent time
intervals, and are expected to be less dependent on the specific environment,
thus simplifying the calibration of the motion detection system during
the system warm-up.

\textbf{RSS: average and standard deviation}. They corresponds to
the first and second moment extracted from $T$ consecutive RSS obtained
from the corresponding PPDU frames. As analyzed below and also verified
in \cite{patwari1}, we found out that the variance is way more accurate
and sensitive than the average value. Instead of using the RSS average
value (in dB) at frame $t$, namely $\mathrm{RSS}(t)$, we replace
it with the relative measure $r(t)$ that considers the temporal difference
of the RSS observed over two adjacent WiFi frames, as follows: $r(t)=\sum_{i=t-T}^{t}\frac{\mathrm{RSSI}(i)}{\mathrm{RSSI}(i-1)}$. 

\textbf{CSI: time and frequency domain correlation}. This feature
exploits the correlation of CSI amplitude $\mathbf{Y}(t)$ in both
time and frequency domains \cite{16}. Time and frequency domain correlation
decreases in the presence of moving entities \cite{16}: to leverage
the correlation in both domains, we adopt the Motion Indicator term
as defined in \cite{16} (Algorithm 1).

\textbf{CSI: phase deviation. }It monitors the fluctuations in the
phase information as induced by body movements. In particular we use
the short-term averaged variance ratio metric defined for each antenna
$m$ as: $\mathrm{SVR}_{m}=\frac{1}{S}\sum_{s=1}^{S}\frac{\mathrm{std_{\text{[\ensuremath{\mathit{t,t+T}}]}}}\widehat{\phi}_{s,m}(t)}{\mathrm{std_{\text{[\ensuremath{\mathit{t-T,t}}]}}}\widehat{\phi}_{s,m}(t)}\cdot\frac{\sum_{i=t}^{t+T}\widehat{\phi}_{s,m}(t)}{\sum_{i=t-T}^{t}\widehat{\phi}_{s,m}(t)}$
, namely the ratio between the normalized (w.r.t. average) phase deviation
observed in the time interval $[t,t+T]$ and the one observed in $[t-T,t]$.

\begin{figure}
	\centering
	\includegraphics[width=0.9\columnwidth]{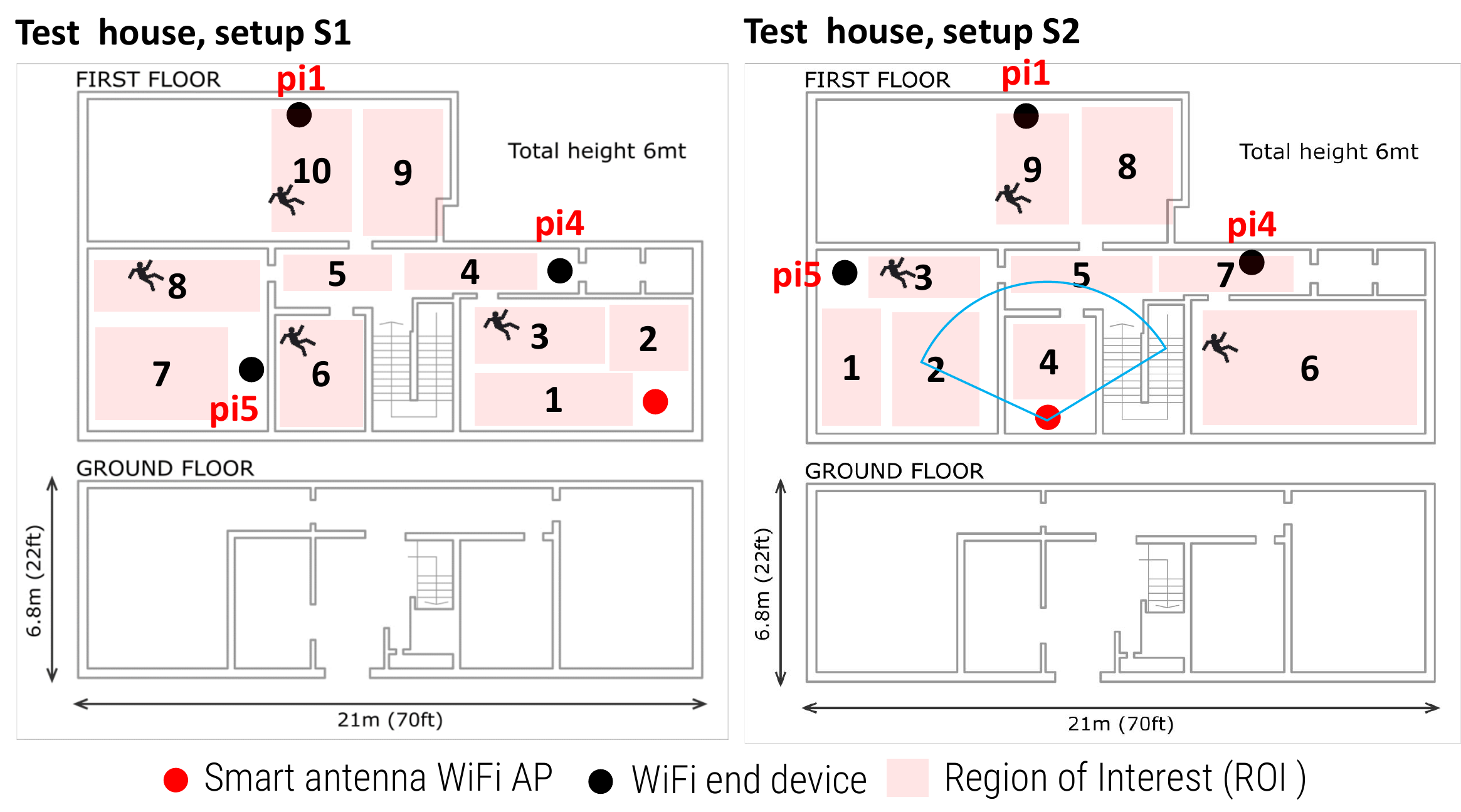}
	\caption{Test house environment: test house setup \#1 and \#2 (floor map), WiFi AP and end device locations (3 WiFi end devices: pi1, pi4, pi5), and corresponding monitored region of interest $1-10$.}
	\label{example}
\end{figure}

Evaluation of the aforementioned features with respect to motion detection
is based on Support Vector Machine (SVM) techniques. In particular,
the SVM weights are trained to classify body motions using data collected
in different environments and AP deployments with respect to the validation
set (see Sect. III). In the following section, we also compare the
accuracy of each individual feature to highlight their advantages/limitations.


\begin{table}
	\centering
	\includegraphics[width=0.9\columnwidth]{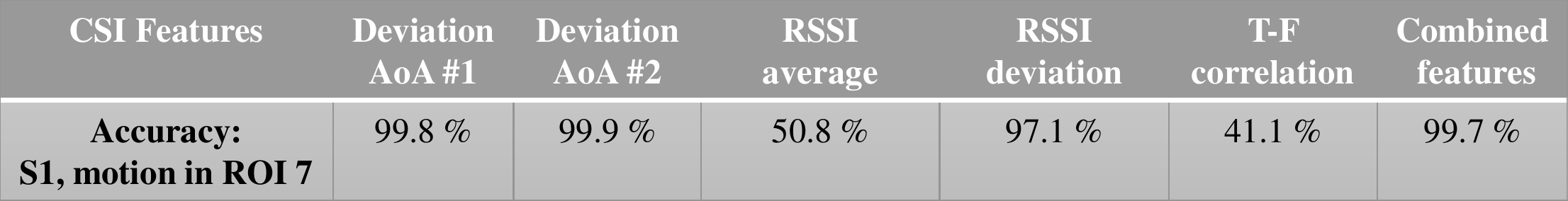}
	\caption{Comparison of CSI features and accuracy for the selected setup S1.}
	\label{features}
\end{table}

\begin{table}
	\centering
	\includegraphics[width=0.8\columnwidth]{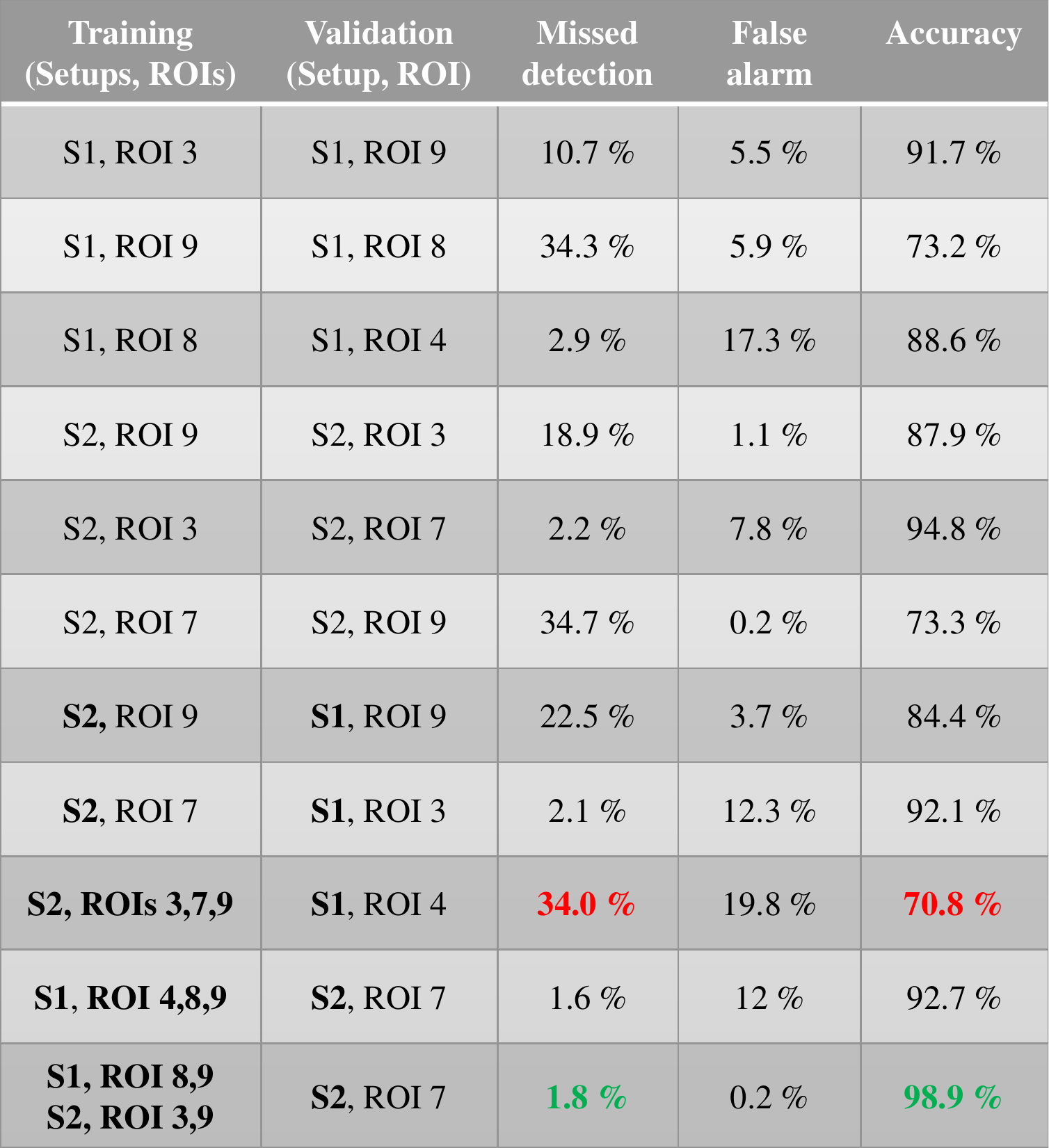}
	\caption{Summary of motion detection performance (missed detection, false alarm probabilities and average accuracy) for combined features and different training/validation setups.}
	\label{results}
\end{table}

\section{Case study in a test house environment}

\label{sec:Motion-discrimination-via}In the proposed case study,
depicted in Fig. \ref{example}, we deployed $N=3$ unmodified WiFi
devices (namely, $\mathrm{p}1,\mathrm{p}4,$ and $\mathrm{p}5$) equipped
with $M_{T}=1$ antennas each and one WiFi AP gateway with $M_{R}=4$
smart antennas \cite{reconfigurable,patent} inside an instrumented
smart-space environment. The antenna module has inter-element spacing
of $D=2.52$ cm at $5.745$ GHz while the AP device CPU is based on
the IPQ8072A Quad Core ARM Cortex A53 System on Chip (SoC) working
at 2.2GHz that supports WiFi 11.ax communications. HW specifications
and data sheets are available online \cite{adant}. The proposed application
case study is meant to design a system that provides augmented sensing
functions for real-time motion detection of a subject located in different
areas, namely the $10$ Region Of Interest (ROIs) indicated in Fig.
\ref{example} for both setups S1 and S2. Setups are characterized
by different locations of the WiFi end device pi5 and the WiFi AP
gateway. To assess the robustness of the detection process, the approach
we followed is to train the system by collecting data from one of
the two setups and then validate the accuracy of the motion detection
system using data from the other setup. With respect to data collection
and processing, the AP serves as an edge node not only for CSI data
collection, but also for AoA and feature processing. In the proposed
settings, the evaluation of the features through SVM is implemented
on a remote PC: the CSI features are moved/transmitted by using JSON
(Java Script Object Notation) serialization and the MQTT publisher/subscriber
transport service \cite{pred}.

In the Tab. \ref{features}, we compare the motion detection accuracy
using selected CSI features as defined in Sect. \ref{beams}. In particular,
the AoA estimation is based on the algorithm described in Sect. II:
we monitor $L=2$ dominant paths, while a motion detection indicator
is issued when a change is detected. Monitoring the dominant paths
$\ell=1,2$ (AoA\#1, AoA\#2) separately gives much higher accuracy
than average RSS and time-frequency correlation. The optimized pool
of features should therefore include both AoAs and RSS standard deviation.
For the selected training and validation setup combinations, in Tab.
\ref{results}, we analyzed the missed detection, the false alarm
probability and the average motion detection accuracy. For each considered
case, we reported the corresponding setup (S1 or S2, or both) and
the ROIs occupied by the subject when training the SVM tool and validating
the motion detection performance, respectively. As shown in this table,
training the SVM model for the same setup (S) used during the system
validation, but different ROIs, leads to high accuracy, as expected
($97$\% on average). On the other hand, training and validating over
different setups and ROIs might lead to significant accuracy drops,
down to $70$\% in some cases. For example, training in ROI 3 (setup
S1) and validating motion performance in ROI 4 gives low accuracy
which is coherent with the position in which the movement has been
made with respect to the WiFi path between the end device and the
WiFi AP. For such cases, it is advisable to train the model using
multiple ROIs, \emph{i.e.} multiple rooms, to enrich the number of
examples in the training set. Finally, training the SVM model using
examples obtained from different WiFi AP deployments (setups S1 and
S2) gives the best accuracy performance ($98$\%).

\section{Concluding remarks}

The paper introduced the use of the beam-steering antenna technology
for passive sensing and detection of body motion events using ambient
WiFi signals. Pattern reconfigurable antennas can channelize the radiation
energy to improve the coverage over selected areas and are thus helpful
for environment-aware joint sensing and communication applications.
A model for beam-space processing has been proposed and applied to
estimate the dominant propagation paths (AoA) using a PCA-based modified
MUSIC algorithm. Validation tests have been carried out inside a smart
home environment with the purpose of verifying the robustness of the
motion detection system inside a representative residential WiFi network.
As future development, the possibility of controlling the antenna
steering process on a physical frame basis will be investigated as
a new opportunity for improving AoA monitoring for sensing. 

\section{Acknowledgements}
This research work has been supported by the CHIST-ERA EU project RadioSense (Wireless Big-Data Augmented Smart Industry) under grant CHIST-ERA-17-BDSI-005.

\end{document}